\documentclass[aps,pra,twocolumn,superscriptaddress,showpacs,10pt]{revtex4-2}
\usepackage{amsmath}
\usepackage[dvips,xdvi]{graphicx}
\usepackage{amssymb}
\usepackage{epsfig}
\usepackage{xcolor}
\usepackage{physics}
\usepackage{soul,color}
\usepackage[colorlinks=true,citecolor=blue,urlcolor=magenta,linkcolor=blue]{hyperref}
\usepackage{microtype}

\newcommand{\Rev}[1]{{\color{black} #1}}

\begin{document}
	\title{Control of dynamical phase transitions and non-ergodic relaxation via spinor phases}
	\author{J. O. Austin-Harris}
	\author{P. Sigdel}
	\author{C. Binegar}
    \affiliation{Department of Physics, Oklahoma State University, Stillwater, Oklahoma 74078, USA}
	\author{S. E. Begg}
    \affiliation{Department of Physics, Oklahoma State University, Stillwater, Oklahoma 74078, USA}\affiliation{Department of Physics, The University of Texas at Dallas, Richardson, Texas 75080, USA}
	\author{T. Bilitewski}
	\email{thomas.bilitewski@okstate.edu}
	\author{Y. Liu}
	\email{yingmei.liu@okstate.edu}
	\affiliation{Department of Physics, Oklahoma State University, Stillwater, Oklahoma 74078, USA}
	\date{\today}
    
	\begin{abstract}
Utilizing ultracold spinor gases as large-scale, many-body quantum simulation platforms, we establish a toolbox for the precise control, characterization, and detection of nonequilibrium dynamics via internal spinor phases. We develop a method to extract the phase evolution from the observed spin population dynamics, allowing us to define an order parameter that sharply identifies dynamical phase transitions over a wide range of conditions. This work also demonstrates a technique for inferring spin-dependent interactions from a single experimental time trace, in contrast to the standard approach that requires mapping a cross section of the phase diagram, with immediate applications to systems experiencing complex time-dependent interactions. Additionally, we demonstrate experimental access to and control over non-ergodic relaxation dynamics, where states of similar energy in the (nominally) thermal region of the energy spectrum retain a dependence on the initial state, via the manipulation of spinor phases, enabling the study of non-ergodic thermalization dynamics connected to quantum scarring. 
	\end{abstract}
	\maketitle

    Ultracold spinor gases, highly controllable quantum systems that possess a spin degree of freedom with all-to-all spin interactions and a well-studied phase space, are quantum simulators ideal for studying a wide variety of nonequilibrium phenomena including quantum scars, quantum many-body scars, and dynamical phase transitions (DPTs)~\cite{Stamper2013,Ueda2012,Kronjager2006,Yingmei2009,Pechkis2013,Zhang2005,Chang2005, Austin3,Zach1,Black2007,Lichao2014,Lichao2015,Zhang2005,Chang2005,Jiang2014,He2015,Gerbier2021,Zach2,Chen2019,Austin1,Austin2,Austin4,Cosmo2026,Jiang2016,Dag2024,Dag2025,Guan2021,Zhou2023,Marino2022}. Both quantum scarring \cite{Heller1984, Turner2018, Pilatowsky2021, Serbyn2021, Hummel2023, Su2023, Zhang2023, Sinha2024,Pizzi_2025} and DPTs \cite{PhysRevLett.110.135704,Heyl_2018,PhysRevLett.126.040602} have attracted attention due to their fundamental importance to our understanding of quantum many-body equilibrium and non-equilibrium physics \cite{Polkovnikov2011, Alessio2016, Deutsch2018, Ueda2020}, as well as due to their potential to advance quantum technology. In particular, quantum scarring has promising applications in quantum transport and quantum metrology~\cite{Dooley2023,bluvstein2022,Turner2018,Serbyn2021}, while DPTs have been suggested as pathways to quantum-enhanced sensing and the generation of entanglement~\cite{Marino2022,Guan2021,Zhou2023}.
    The spinor physics underlying these phenomena is deceptively simple and can often be described using just two types of observables: spin populations and spinor phases. Despite the vital importance of the spinor phases in characterizing both ground-state and excited-state phase diagrams as well as dynamics~\cite{Chapman2005PRL,Sengstock2006,Yi2003,Zhang2005,Chang2005,Pechkis2013,Austin3}, experimental studies of spinor physics have thus far largely relied on the observation of spin population dynamics due to the technical challenge of directly measuring spinor phases~\cite{Stamper2013,Pechkis2013,Zhang2005,Chang2005,Kronjager2006,Black2007,Yingmei2009,Lichao2014,Jiang2014,Lichao2015,Zach1,Ueda2012,Austin3,Jiang2014,He2015}. This population-only approach, however, can obscure understanding of the spinor physics and important connections to other physical systems~\cite{Chang2005, Zach1, Austin3}. In particular, experimental studies of DPTs in spinor gases have been hindered by the lack of identification of suitable order parameters~\cite{Yingmei2019,Marino2022,Austin3,Tian2020,Feldmann2021}.

    Here, we show that spinor phases are a vital tool for state preparation, and enable control and characterization of nonequilibrium dynamics, including diagnosing DPTs in static and driven lattices, and further demonstrate phase-dependent non-thermal long-time behavior in the quantum many-body dynamics. We experimentally demonstrate that an order parameter $\beta(\theta)$, based on the relative phase $\theta$ among all spin components, is capable of sharply distinguishing a DPT between the interaction-dominated regime and the Zeeman-dominated regime. Observables directly based on spin population measurements, such as the period or center of spin oscillations, are not capable of identifying what regime of the phase diagram the system is in without comparing to the theoretical phase diagram and are therefore insufficient for diagnosing DPTs~\cite{Austin3,Yingmei2019,Qingze2025}, in particular in scenarios where there are no \textit{a priori} established theory results. Conversely, spinor-phase-based observables, as developed here, can provide a more rigorous method of directly obtaining the dynamical phase diagram and characterizing DPTs~\cite{Chang2005}. We also develop a method for determining spin-dependent interactions from a single time trace, in contrast to the standard method that requires deliberately mapping a cross section of the phase diagram, with immediate applications to systems experiencing complex time-dependent interactions~\cite{Black2007,Lichao2014,Lichao2015,Zach1,Austin3}. Finally, we show that control over spinor phases is critical for state preparation in $U(1)$ symmetry broken systems. In this system, we demonstrate that states of similar energy in the (nominally) thermal region of the energy spectrum can display non-thermal values at late times, retaining a dependence on two relative spinor phases, $\theta$ and $\eta$, where $\eta$ is the relative phase among components of nonzero spin. Together, this work establishes experimental control over spinor phases as a powerful tool for probing and controlling nonequilibrium dynamics.

    \begin{figure*}[tb]
		\includegraphics[width=176mm]{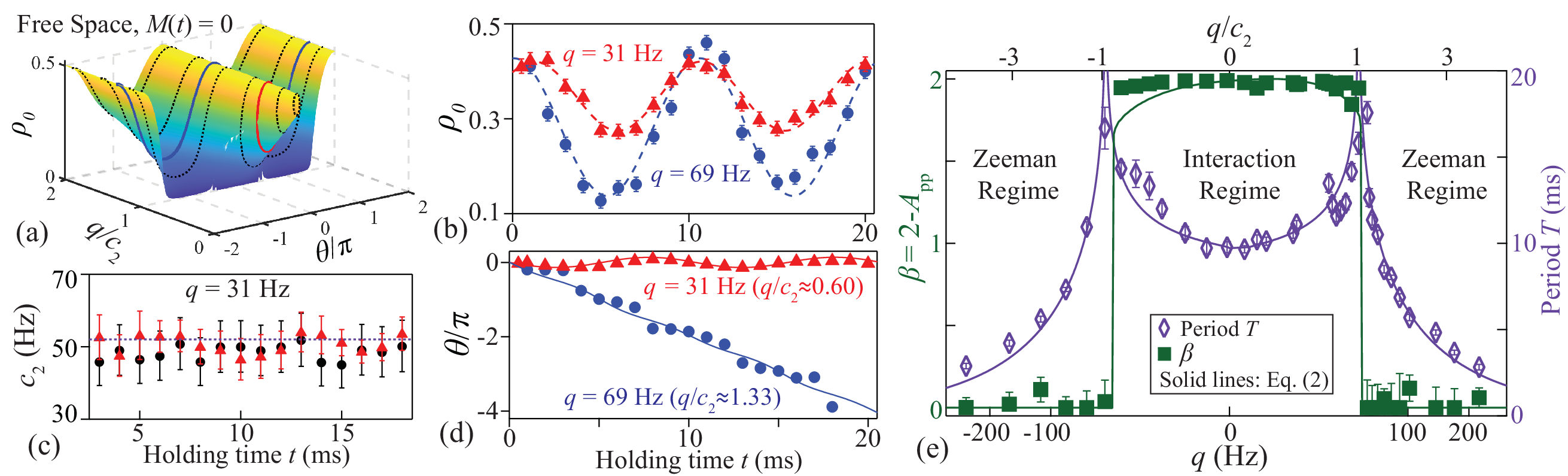}
		\caption{(a) Equal energy contours, derived from Eq.~\eqref{FS_Ham} for the \textit{S}-state, demonstrate the phase diagram consisting of the interaction regime (where $|q/c_2|<1$) and the Zeeman regime (where $|q/c_2|>1$). The red (blue) solid contour marks $q/c_2=0.60$ ($q/c_2=1.33$).
        (b) Triangles (circles) display $\rho_0$ dynamics observed after a quench in $q$ from $41~\mathrm{Hz}$ to $31~\mathrm{Hz}$ ($69~\mathrm{Hz}$) at time $t=0$ from the \textit{S}-state in free space. %
        (c) $c_2$ (triangles) extracted from the $q=31~\mathrm{Hz}$ spin dynamics shown in panel~(b); consistent with values (dotted line) inferred from the observed separatrix shown in panel~(e) and results (circles) derived from the observed atom number and trapping frequencies. %
        (d) $\theta$ extracted from the $\rho_0$ time traces shown in panel (b). %
        (e) Squares (diamonds) display the order parameter $\beta$ ($T$) mapping the experimental phase diagram after a sudden quench in $q$ from $41~\mathrm{Hz}$ to a given value for the initial \textit{S}-state in free space. Purple data points are adapted from Ref.~\cite{Lichao2014}. In panels (b)-(e) solid (dashed) lines are Eq.~\eqref{FS_Ham} predictions (eye-guiding fits).  } \label{FreeSpace}
	\end{figure*}

{\it Experimental Sequence---}%
Each experimental cycle begins with an $F=1$ spinor Bose-Einstein condensate (BEC) of up to $10^5$ sodium atoms in a crossed optical dipole trap.  The desired initial state is then
prepared using a resonant radio-frequency (RF) pulse. We then commence each experimental sequence, in which specific parameters of the system are varied by applying microwave dressing fields, optical lattices, or RF driving fields at the time $t=0$ as described in the following paragraphs (with additional details in Appendix~\ref{A}). The atoms are trapped for a varied hold time $t$ before being released for spin-resolved imaging. For the data presented, all spin states appear to share the same spatial mode, which is supported by the calculated spin healing lengths being larger than the Thomas-Fermi radii, allowing the system to be described with a single spatial-mode approximation (SMA) (see Appendix~\ref{A}).

{\it Model---}%
When the SMA is valid, the spin dynamics of an $F=1$ spinor BEC of $N$ atoms subject to a large static magnetic field along the $z$-axis and a much weaker time-dependent field along the $y$-axis can be described by 
\begin{equation}
	\hat{H}=\frac{c_2}{2N} \mathbf{S}^2+p_B \hat{S}_z+q\sum_i(\hat{s}_{i,z})^2 +p(t)\hat{S}_y,  \label{SMA_Ham}
\end{equation}
with $c_2$ the spin-dependent interaction, $p_B$ ($q$)  the linear (quadratic) Zeeman shift associated with the strong static magnetic field, $p(t)$ the time-variant linear Zeeman interaction induced by the weaker magnetic field, $\mathbf{S}$ the spin operator, and $\hat{S}_y$ and $\hat{S}_z$ the respective components.

A spin-coherent initial state $\lvert \Psi \rangle = \prod_i \sum_{m_F }\psi_{m_F} \hat{a}_{m_F}^{\dagger}\lvert \mathrm{vac}\rangle$ is characterized by three complex amplitudes of the $m_F$ hyperfine spin states $(\psi_1, \psi_0, \psi_{-1})=(\sqrt{\rho_1} e^{i \phi_1}, \sqrt{\rho_0} e^{i\phi_0}, \sqrt{\rho_{-1}} e^{i \phi_{-1}})$ in terms of the fractional populations $\rho_{m_F}$ and phases $\phi_{m_F}$. Here $\hat{a}^{\dagger}_{m_F}$ is the creation operator for a boson in the $m_F$ hyperfine level and $\lvert \mathrm{vac} \rangle$ is the vacuum state. The populations may be written as  $\rho_{\pm1} = (1-\rho_0 \pm M)/2$ in terms of $\rho_0$ and the magnetization $M = \rho_{1}- \rho_{-1}$, and the phases can be parametrized in terms of  $\theta = \phi_1+\phi_{-1}-2\phi_0$ and $\eta=\phi_1-\phi_{-1}$. In total, the state is then fully characterized by $\rho_0$, $M$, $\theta$ and $\eta$. %

Broadly, prior experimental studies of spinor gases have mainly focused on population dynamics~\cite{Black2007,Yingmei2009,Yingmei2019,Austin1,Austin2,Austin3,Austin4,Chen2019,Zach1,Zach2,Stamper2013,Ueda2012,Jiang2014,Lichao2014,Lichao2015,He2015,Gerbier2021,Pechkis2013,Kronjager2006,Chang2005,Tian2020,Cosmo2026}, tracking $\rho_0$ only or $\rho_0$ and $M$. In contrast, in this work we control the phases and populations of spinor BECs and highlight how (i) extraction of the information of spinor phases, for example $\theta$, provides significant advantages especially in identifying DPTs, and (ii) scenarios in which both phases $\theta$ and $\eta$ significantly affect the dynamics and determine the relaxation values of spinor BECs.

{\it Spinor phases \& dynamical phase transitions---}%
In the non-driven case, where $p(t)=0$, the system has $U(1)$ symmetry around $\hat{S}_z$ and the spin dynamics can be described by $\rho_{0}$, the magnetization $M$, and a single relative phase $\theta$ by the mean-field Hamiltonian~\cite{Stamper2013,Pechkis2013,Zhang2005,Chang2005,Kronjager2006,Black2007,Yingmei2009,Lichao2014,Jiang2014,Lichao2015,Zach1,Ueda2012, Austin3}:
	\begin{equation}
		H/h=c_2\rho_0[(1-\rho_0)+\sqrt{(1-\rho_0)^2-M^2}\cos(\theta)]+q(1-\rho_0),\label{FS_Ham}
	\end{equation} 
	with the associated equations of motion~\cite{Zhang2005,Chang2005,Lichao2014}
\begin{align}
    \pdv{\rho_0}{t}&=\frac{-2}{\hbar}\pdv{H}{\theta}=\frac{c_2}{\pi}\rho_0\sqrt{(1-\rho_0)^2-M^2}\sin(\theta)\label{RhoEqOfMotion}
\end{align} 
\begin{align}
		\pdv{\theta}{t}&=\phantom{-}\frac{2}{\hbar}\pdv{H}{\rho_0}=\frac{c_2}{\pi}\frac{(1-\rho_0)(1-2\rho_0)-M^2}{\sqrt{(1-\rho_0)^2-M^2}}\cos(\theta)\notag\\
		&\qquad \qquad \qquad +\frac{c_2}{\pi}(1-2\rho_0)-\frac{q}{\pi}.\label{ThetaEqOfMotion}
\end{align} 
As shown in Fig.~\ref{FreeSpace}(a), the resulting spin dynamics can be categorized into two regimes: the Zeeman regime where $\theta$ is unbounded (see the blue contour) and the interaction regime where $\theta$ is bounded (see the red contour). These regimes are separated by a separatrix in phase space whose precise location depends on the initial state. For the initial \textit{S}-state where $\rho_0(0)\approx0.45$, $M(0)=0$, and $\theta(0)=0$, the separatrix is at $c_2/q\approx\pm 1$ (see Fig.~\ref{FreeSpace}).

Experimental spin population time traces also carry information about the interactions and relative spinor phase $\theta$, which can be extracted from the dynamics using Eqs.~\eqref{FS_Ham}-\eqref{ThetaEqOfMotion}. Typical examples of $\rho_0$ dynamics observed in free space after a sudden quench in $q$ via the application of a microwave dressing field are displayed in Fig.~\ref{FreeSpace}(b). For certain parameter regimes, $c_2$ can be reliably extracted directly from the observed spin dynamics and Eqs.~\eqref{RhoEqOfMotion}-\eqref{ThetaEqOfMotion} as demonstrated by the triangles in Fig.~\ref{FreeSpace}(c) (see Appendix~\ref{ThetaExt}). This new method enables $c_2$ to be experimentally confirmed from a single time trace rather than the many time traces needed to fully map a cross section of the phase diagram and determine the separatrix location to measure $c_2$~\cite{Black2007,Lichao2014,Lichao2015,Zach1,Austin3}. In simple systems, e.g., in free space, this new method returns results for $c_2$ consistent within error with other methods, for example the value (dotted line) inferred from the detected separatrix location or the results (circles) based on the observed atom number and trapping potentials as shown in Fig.~\ref{FreeSpace}(c). However, in more complex systems subject to violent spatial dynamics or unknown time-variant trapping potentials, e.g., a moving lattice system, our data in Ref.~\cite{FutureWork} indicate that the new method may be necessary to reliably extract $c_2$. Once $c_2$ is known, $\theta$ can also be extracted from the spin dynamics using Eqs.~\eqref{RhoEqOfMotion} and \eqref{ThetaEqOfMotion} (see Appendix~\ref{ThetaExt}), as shown in Fig.~\ref{FreeSpace}(d). For these datasets, the evolution of $\theta$ is markedly dissimilar (see Fig.~\ref{FreeSpace}(d)), despite the observed $\rho_0$ dynamics being superficially similar (see Fig.~\ref{FreeSpace}(b)). Specifically, in the interaction regime, $\theta$ oscillates between bounds as shown by the $q/c_2\approx0.60$ data (triangles), while in the Zeeman regime, $\theta$ evolves monotonically with time as shown by the $q/c_2\approx1.33$ data (circles) consistent with theoretical predictions (see Fig.~\ref{FreeSpace}(d)).

\begin{figure}[t]
		\includegraphics[width=86mm]{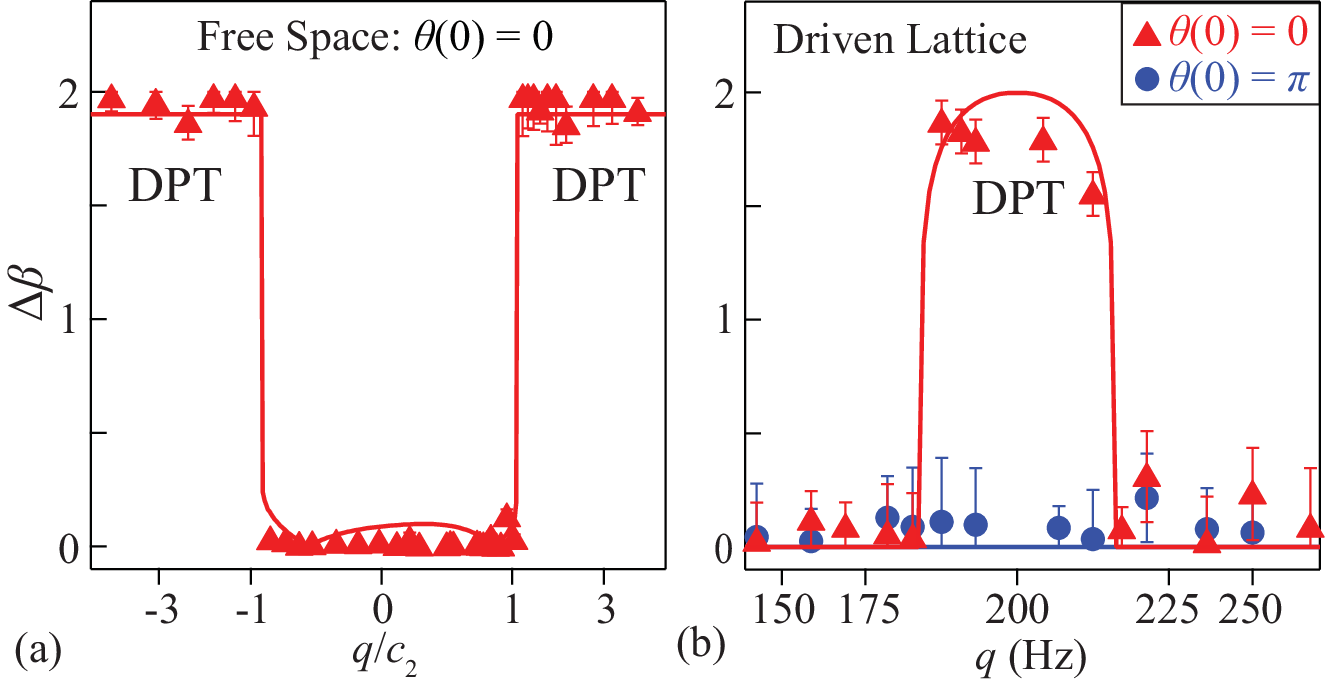}
		\caption{(a) Markers display $\Delta\beta$, the change in $\beta$ after a sudden quench in $q$, for the data shown in Fig.~\ref{FreeSpace}(e), versus $q/c_2$. No DPT occurs when $\Delta\beta\sim 0$. (b) Triangles (circles) display $\Delta\beta$ after the application of a 1D lattice whose depth is driven sinusoidally at a frequency $f=400~\mathrm{Hz}$ between 0 and $5E_R$ starting from the initial state of $\rho_0(0)\approx0.45$, $M=0$, and $\theta(0)=0$ ($\theta(0)=\pi$). Solid lines are SMA predictions.} \label{OrderParam}
    \end{figure}

While information that is readily apparent in spin population dynamics, such as spin oscillation period or average $\rho_0$, is insufficient to serve as order parameters to map out dynamical phase diagrams, information about $\theta$, hidden in the dynamics, can be extracted using Eq.~\eqref{RhoEqOfMotion} and Eq.~\eqref{ThetaEqOfMotion} (see Appendix~\ref{ThetaExt}) to uniquely identify the regime of the phase diagram. Phase-based order parameters have been experimentally underexplored due to the technical challenge of directly obtaining information about $\theta$~\cite{Austin3, Yingmei2009, Qingze2025}. Here, we demonstrate an order parameter $\beta=2-A_{\mathrm{pp}}$ via indirect measurements of $\theta$ from spin dynamics observed using standard absorption imaging (see Fig.~\ref{FreeSpace}(d)). Here $A_{\mathrm{pp}}=\max[\cos(\theta/2)]-\min[\cos(\theta/2)]$ is the peak-to-peak value of an experimental $\cos(\theta/2)$ oscillation. Our data in Fig.~\ref{FreeSpace}(e) show an advantage of the order parameter $\beta$: a sharp phase transition is identified by $\beta$ as the system passes between the interaction regime, an ordered regime in which $\beta$ is nonzero (and close to two), and the Zeeman regime, a disordered regime in which $\beta$ is zero. This experimental result is well described by the SMA model (see the green line in Fig.~\ref{FreeSpace}(e)). In contrast, observables directly based on spin populations, such as the spin oscillation period (diamonds in Fig.~\ref{FreeSpace}(e)), do not clearly distinguish between the regimes because  \textit{a priori} knowledge of the phase diagram is needed to determine what regime the system is in for a given set of parameters. Additionally, our data suggest that $\Delta\beta=\abs{\beta(t>0)-\beta(t<0)}$ can characterize DPTs by indicating a nonanalytic change in $\beta$ during a quantum quench. For example, for the data in Fig.~\ref{OrderParam}(a), because the prequench state is in the interaction regime with $\beta\approx1.98(2)$, quenches to the Zeeman regime achieve a DPT reflected by $\Delta\beta \gg 0$ due to the quench-induced sudden change in $\beta$ from near two to zero (see the $\abs{q/c_2}>1$ regions in Fig.~\ref{OrderParam}(a))~\cite{Marino2022,Robert2021,Austin3}. Conversely, if $\Delta\beta$ is approximately zero it indicates that no DPT occurs as the system remains in the same regime of the phase diagram after the quench (see the $\abs{q/c_2}<1$ region in Fig.~\ref{OrderParam}(a)). 

   \begin{figure*}[tb]
    	\includegraphics[width=176mm]{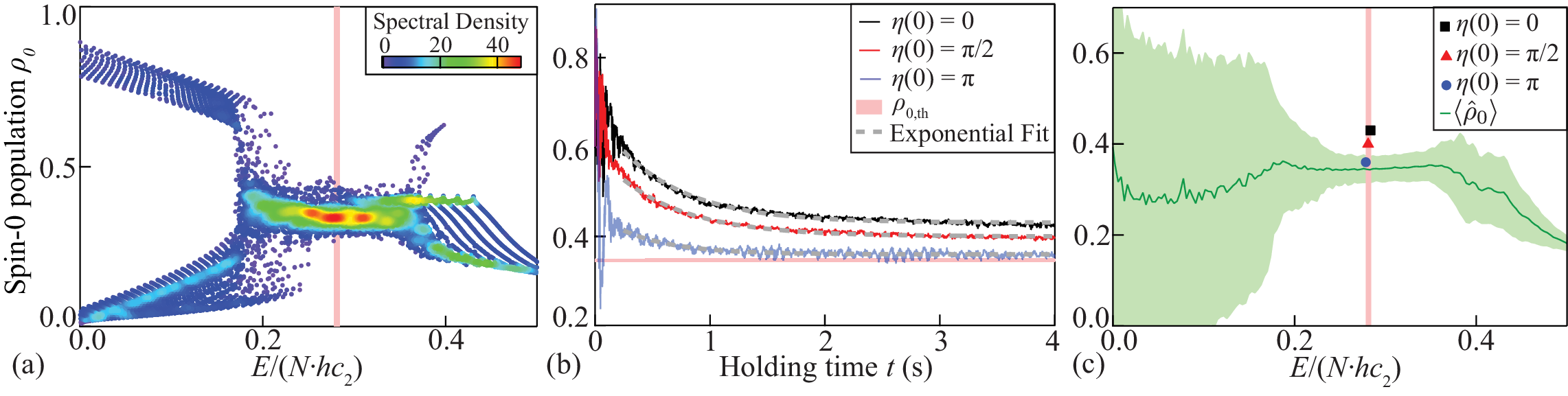}
		\caption{%
        (a) Expectation values of $\rho_0$ in eigenstates versus energy. Pink vertical lines show the energy range of \textit{U}-states, a group of spin-coherent initial states with $\rho_0(0)=0.6$, $M(0)=0$, and $\theta(0)=\pi$ with varying $\eta(0)$. (b) Time-traces of $\rho_0$ for \textit{U}-states with $\eta(0) = 0$, $\pi/2$, $\pi$. The shaded region marks the range of ETH expectation values at the corresponding energy region (see the pink shaded region in panel~(a)). \Rev{Grey dashed lines are exponential fits to the data. (c) The green line shows the microcanonical expectation value $\langle \hat{\rho}_0 \rangle$ (see text), while the green shaded region indicates one standard deviation around it. The pink shaded region shows the energy range of \textit{U}-states. The three markers show the extrapolated values of $\rho_0(t\rightarrow \infty)$ extracted from the fitting curves in panel (b) for $\eta=0,$ $\pi/2,$ and $\pi$.}  All calculations in this figure were performed for $q=40~\mathrm{Hz}$ and $p\approx 8~\mathrm{Hz}$.
         \label{Eta}}
	\end{figure*} 

The study of DPTs using $\Delta\beta$ can be extended to more complicated systems. For example, one-dimensional (1D) sinusoidally driven optical lattices allow engineering the phase diagram of spinor gases, tuning multiple parameters including the effective $\theta(0)$ and generating additional separatrixes at multiples of half the driving frequency $f$~\cite{Austin3}. These separatrixes can be described using a model identical to the standard SMA model (Eq.~\eqref{FS_Ham}) where $q$ and $\theta$ are replaced with effective quantities and the strengths of the spin-changing and spin-preserving collisions can be independently tuned~\cite{Austin3,Qingze2025}. Figure~\ref{OrderParam}(b) shows typical examples characterized by $\Delta\beta$ for two different initial states in a driven-lattice system in which the depth of a 1D lattice is sinusoidally driven between 0 and $5E_R$ (with $E_R$ the recoil energy) at $f=400~\mathrm{Hz}$, inducing effective quenches of both the quadratic Zeeman energy and interactions. For the initial \textit{S}-state where $\theta(0)=0$ (red triangles), the system undergoes a DPT in the region identified by $\Delta\beta\gg0$ (see the vicinity of $q\sim f/2=200~\mathrm{Hz}$ in Fig.~\ref{OrderParam}(b)). Meanwhile, for the other initial state where $\theta(0)=\pi$ (blue circles in Fig.~\ref{OrderParam}(b)), we observe that $\Delta\beta$ remains close to zero for all $q$ indicating that, for this initial state, the system does not undergo a DPT during the quenches. These observations agree well with the SMA model (see the solid lines in Fig.~\ref{OrderParam}(b)).

    {\it Non-thermal phase-dependent relaxation dynamics---}%
    The phase $\eta$ becomes relevant in the absence of $U$(1) symmetry, which we break by introducing a time-dependent field $p(t)$ resulting in magnetization dynamics and full dependence on all spinor phases. A particularly interesting example is adding a weak near-resonant driving field, i.e., $p(t)=p(\sin(\omega_{+}t)+\sin(\omega_{-}t))$ where $\omega_\pm=2\pi(p_B\pm q)$ for $q\gg p$. This model has been shown to host both quantum many-body scars (QMBS) and quantum scars induced by an underlying unstable periodic orbit (UPO), enabling the study of the connections between the two distinct types of scarring~\cite{Austin4,Dag2024,Dag2025}.

   The mean-field Hamiltonian for this system is given by
	\begin{align}
		H/h &= c_2\rho_0[(1-\rho_0)+\sqrt{(1-\rho_0)^2-M^2}\cos(\theta)] \notag\\
        & + q(1-\rho_0) + p_B M \label{eq:H_mf_drive}\\
        &+p(t) \sqrt{8\rho_0} \left(\sqrt{\rho_{1}} \sin(\frac{\theta+\eta}{2}) - \sqrt{\rho_{-1}} \sin(\frac{\theta-\eta}{2})\right) \notag
	\end{align}
    where the dynamics now involves all of $\rho_0$, $M$, $\theta$ and $\eta$. 
    
     We first consider the spectral properties of the corresponding Floquet Hamiltonian (see Appendix~\ref{theory}). Fig.~\ref{Eta}(a) shows the eigenstate expectation values of $\rho_0$ versus eigenenergy $E$. This shows a fully thermal region for $0.22 \lesssim E/(N h c_2) \lesssim 0.38$ in the middle of the spectrum, as well as athermal regular regions at low- and high-energy at the edges of the spectrum. 
     Additionally, for $0.13\lesssim E/(N h c_2) \lesssim 0.22$ there is a region with QMBS coexisting with thermal states. In prior work \cite{Austin4}, we showed that this can lead to the absence of equilibration at late times for initial states overlapping with these states. 
     Here, we study the dependence on the initial phase $\eta(0)$. We select a group of spin-coherent initial states with $\rho_0(0)=0.6$, $M(0)=0$, and $\theta(0)=\pi$, which we refer to as \textit{U}-states. Crucially, the range of resulting energies of \textit{U}-states is almost independent of $\eta(0)$, as marked by the pink shaded region in Fig.~\ref{Eta}(a) \footnote{For the chosen initial states, with $\theta(0)=\pi$ and $M(0)=0$, the mean-field initial state energy is completely independent of $\eta(0)$ (see Eq.~\eqref{eq:H_mf_drive}). The residual $\eta$ dependence appears only from higher-order Floquet corrections}. For $\rho_0(0)$ between about 0.52 and 0.63 the initial state energy for all $\eta(0)$ then lies within the thermal region of the spectrum without QMBS (see Fig.~\ref{Eta}(a)).  %
    
    Fig.~\ref{Eta}(b) shows the exact spin-0 population dynamics $\rho_0(t)$ starting from \textit{U}-states with different $\eta(0)$ values:  $\eta(0)=0$, $\pi/2$, and $\pi$. We observe that while all \textit{U}-states do relax at long times \Rev{(see Fig.~\ref{Eta}(b))}, they do not generally relax to the expected micro-canonical eigenstate thermalization hypothesis (ETH) prediction~\cite{Deutsch1991,Srednicki1994,Deutsch2018}, indicated by the pink shaded region in Fig.~\ref{Eta}(b). Indeed, only for $\eta(0) = \pi$ and $\eta(0)=3\pi$ (not shown) do we observe thermalization to the expected value, whereas states with generic $\eta(0)$ fail to thermalize, instead relaxing to values that strongly depend on the initial phase. %
    \Rev{Figure~\ref{Eta}(c) displays the microcanonical expectation value $\langle \hat{\rho}_0\rangle(E)$ (with shading indicating the microcanonical standard deviation), that all initial states at energy $E$ are expected to thermalize to. By comparing the extrapolated long-time relaxed values of $\rho_0$ extracted from the fitting lines in Fig.~\ref{Eta}(b) to this thermal prediction, we find that they deviate by as much as three standard deviations from the thermal predictions (see the three markers in Fig.~\ref{Eta}(c)).}
    
\begin{figure}[t]
		\includegraphics[width=86mm]{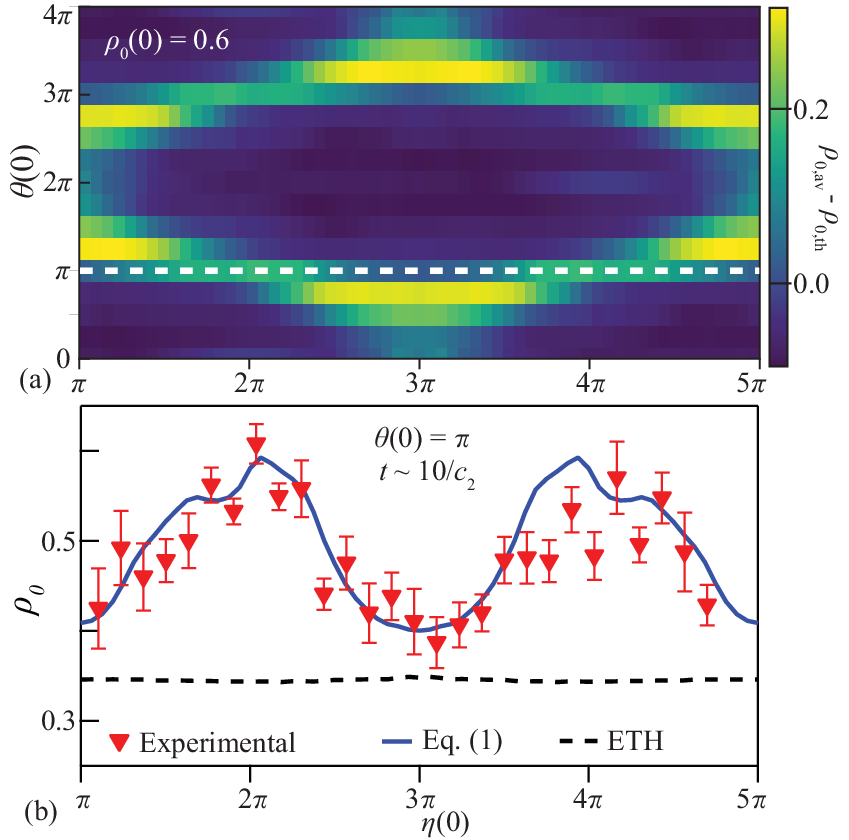}
		\caption{ (a) Deviation of long-time relaxed values at $t\approx1~\mathrm{s}$ from ETH predictions, $\rho_{0,\mathrm{av}}-\rho_{0,\mathrm{th}}$, for spin-coherent states with $\rho_0(0)=0.6$ and $M(0)=0$ versus $ \theta(0)$ and $\eta(0)$. The white dashed line marks \textit{U}-states. (b) Triangles display the equilibrated $\rho_0$, observed at a long holding time of $t=300~\mathrm{ms}\sim 10/c_2$, as a strong function of the initial phase $\eta(0)$ for \textit{U}-states at $q=40~\mathrm{Hz}$ and $p\approx 8~\mathrm{Hz}$ (see Appendix~\ref{A}). The solid (dashed) line is the Eq.~\eqref{SMA_Ham} (ETH) prediction. All theory calculations in this figure were performed for $q=40~\mathrm{Hz}$ and $p\approx 8~\mathrm{Hz}$.} \label{EtaExp}
    \end{figure}

    We systematically demonstrate this deviation of long-term relaxed values from thermal values in Fig.~\ref{EtaExp}(a) for spin-coherent states with $\rho_0(0) = 0.6$ and $M(0)=0$ as a function of both phases $\theta(0)$ and $\eta(0)$. The white dashed line in Fig.~\ref{EtaExp}(a) marks \textit{U}-states, which show a maximal deviation from the thermal values for $\eta(0)=2\pi$ and $4\pi$ and thermal behavior for $\eta(0) =\pi$ and $3\pi$. %
    These theory predictions are confirmed by our experimental data in Fig.~\ref{EtaExp}(b), which shows the observed $\rho_0(t)$ (triangles) as a function of the initial phase $\eta(0)$ after $t\sim 10/c_2$. This holding time is long enough that the initial transient spin dynamics have settled out but short enough to avoid unmodeled energy dissipation processes and relaxation channels becoming significant (see Ref.~\cite{Austin4}). Because the phase $\eta$ evolves on a timescale governed by the linear Zeeman effect, by varying a very short delay time between the initial state preparation and the application of the driving fields, $\eta(0)$ can be experimentally tuned effectively independently of $\theta(0)$, which evolves on a much longer timescale governed by $q$ and $c_2$~\cite{Austin4}. For initial \textit{U}-states, our data presented in Fig.~\ref{EtaExp}(b) shows excellent agreement with the theoretical simulations (solid lines) consistent with the predicted slow approach to long-time non-thermal values with strong dependence on the initial phase $\eta(0)$. 
    
    Unlike the nonthermalizing dynamics connected to QMBS studied in our prior work~\cite{Austin4}, the behavior studied here is not reflected in the thermal expectation values and is not believed to be connected to QMBS. Microscopically, we can understand the nonthermalizing dynamics studied in Fig.~\ref{Eta} and Fig.~\ref{EtaExp} by the initial state having overlaps with eigenstates in which the $\rho_0$ expectation values are different from the thermally expected value. Surprisingly, this occurs here in a quantum-many-body system for simple spin-coherent initial product states, with energy expectation value fully within the thermal region of the spectrum, and is highly tunable by the phase $\eta(0)$.

    {\it Discussion \& Outlook---}
    Our results establish experimental control over internal spinor phases as a powerful tool for probing and characterizing quantum non-equilibrium dynamics in spinor BECs. We develop a method to infer spin-dependent interactions and the relative spinor phase $\theta$ from the observed spin dynamics.  This method allows experimental access to new order parameters that sharply identify DPTs in spinor gases over a wide range of conditions. 
    This includes cases where interaction coefficients and the phase diagram are \textit{a priori} unknown or difficult to resolve, such as the driven lattice case~\cite{Austin3,Qingze2025}, and we anticipate extensions to more exotic nonequilibrium protocols such as moving lattices~\cite{FutureWork}.
    Additionally, we experimentally demonstrate that control of the initial spinor phase can be used to induce non-ergodic relaxation dynamics, where states of similar energy in the (nominally) thermal region of the energy spectrum display non-thermal values at long holding times, with a strong dependence on the initial state. Our results mature and advance quantum simulation capabilities in spinor BECs via leveraging full phase control. In turn, this is anticipated to assist quantum computation and sensing protocols, such as for critically enhanced quantum sensing in the proximity of DPTs \cite{Guan2021,Zhou2023},  or state-preparation for quantum metrology and entanglement generation with QMBS states \cite{Dooley2023,bluvstein2022}.   \vspace{-1pc}
	
	\begin{acknowledgments}\vspace{-0.5pc}
		\noindent{\it Acknowledgments --} 
        We acknowledge support from the Noble Foundation and the National Science Foundation (NSF) through Grants PHY-2207777, PHY-2513302, and DGE-2510202. Some of the computing for this project was performed at the High Performance Computing Center at Oklahoma State University supported in part through the NSF grant OAC-1531128. TB acknowledges support by the Air Force Office of Scientific Research under award number FA9550-25-1-0340. S.E.B. acknowledges support from the NSF through grant OSI-2228725.
	\end{acknowledgments}
\vspace{-1pc}
\appendix
    \section{Experimental Details}\label{A}\vspace{-1pc}
    In each experimental sequence described in this work, we generate a $F=1$ spinor Bose-Einstein condensate (BEC) of up to $10^5$ sodium ($^{23}$Na) atoms in a crossed optical dipole trap (ODT). The desired initial state is prepared using a short resonant radio-frequency (RF) pulse. The atoms are then held in the ODT during each experimental cycle while microwave dressing fields, optical lattices, or RF-driving fields are applied to vary specific parameters of the system. After a varied holding time $t$, the atoms are abruptly released from all trapping potentials for time-of-flight expansion followed by spin resolved imaging. Details specific to each experimental sequence are described in the following subsections.    

    \subsection{Spinor BECs subject to microwave dressing fields} \vspace{-1pc}
    For the experiments described in Figs.~\ref{FreeSpace} and \ref{OrderParam}(a), we apply microwave dressing fields immediately after the initial state preparation. These microwave dressing fields are produced using a $\sigma_{\pm}$ polarized microwave pulse detuned from the $\ket{F=1, m_F=0}\rightarrow \ket{F=2,m_F=0}$ transition to quench the quadratic Zeeman shift from its initial value $q_i=41~\mathrm{Hz}$ to a final value $q$ after the quench as detailed in our prior work~\cite{Lichao2014}. For typical condensates studied using this experimental sequence, the spin healing length is around $13~\mu\mathrm{m}$ and the Thomas-Fermi radii are approximately $(6, 6, 4)~\mu\mathrm{m}$.   
    
    \subsection{Spinor BECs in driven optical lattices}\vspace{-1pc}
    For the experiments described in Fig.~\ref{OrderParam}(b), the atoms are adiabatically loaded into a one-dimensional optical lattice with lattice spacing of $532~\mathrm{nm}$ before the initial state is prepared at a lattice depth of $5E_R$ ($0E_R$) for the effective $\theta(0)=0$ $(\theta(0)=\pi)$ data. Here $E_R=h^2 /(2M_{\mathrm{Na}}\lambda^2)$ is the recoil energy, with $M_{\mathrm{Na}}$ being the atomic mass of sodium and $\lambda=1064~\mathrm{nm}$ being the wavelength of the lattice beam. Immediately after the state preparation the lattice depth is sinusoidally varied between 0 and $5E_R$ until the atoms are released for imaging as detailed in our prior work~\cite{Austin3}. For typical condensates studied using this experimental sequence, the spin-healing length is approximately $13~\mu\mathrm{m}$ and the Thomas-Fermi radii are approximately $(9, 9, 7)~\mu\mathrm{m}$. 
    
    The dynamics of this system feature additional separatrixes at integer $j$ multiples of half the driving frequency $f$ that can be described using the following mean-field Hamiltonian under a dynamical SMA:~\cite{Austin3,Qingze2025}:
	\begin{align}
		\hat{H}_{\mathrm{mf},j}/h=&q_{\mathrm{eff},j}(1-\rho_0)+\mathcal{G}_0\rho_0(1-\rho_0)\notag\\
		&+\frac{\mathcal{G}_j}{2}\rho_0\sqrt{(1-\rho_0)^2-M^2} \cos(\theta_{\mathrm{eff},j}).\label{DL_Ham}
	\end{align}
	which is identical to the standard SMA model (Eq.~\eqref{FS_Ham}), except $q$ and $\theta$ are replaced with effective quantities $q_{\mathrm{eff},j}=q-jf/2$ and $\theta_{\mathrm{eff},j}$ and the strengths of the spin-changing and spin-preserving collisions, $\mathcal{G}_j$ and $\mathcal{G}_0$ respectively, can be independently tuned. By changing the characteristics of the driven lattice, all the parameters that determine spinor physics can be controlled, including the effective $\theta$, which can be used to control whether a given quench induces a DPT~\cite{Austin3}. The equations of motion for this model~\cite{Austin3,Qingze2025},	    
\begin{align}
    \pdv{\rho_0}{t}&\!=\frac{-2}{\hbar}\pdv{H_{\mathrm{mf},j}}{\theta_{\mathrm{eff},j}}\!=\!\frac{\mathcal{G}_j}{2\pi}\rho_0\sqrt{(1\!-\!\rho_0)^2\!-\!M^2}\sin(\theta_{\mathrm{eff},j})
\end{align} 
\begin{align}
		\pdv{\theta_{\mathrm{eff},j}}{t}\!=\phantom{-}\frac{2}{\hbar}\pdv{H_{\mathrm{mf},j}}{\rho_0}\!=&\frac{\mathcal{G}_j}{2\pi}\frac{(1\!-\!\rho_0)(1\!-\!2\rho_0)\!-\!M^2}{\sqrt{(1-\rho_0)^2-M^2}}\cos(\theta_{\mathrm{eff},j})\notag\\
		&+\frac{\mathcal{G}_0}{\pi}(1-2\rho_0)-\frac{q_{\mathrm{eff},j}}{\pi}~,\label{eq:H_DrivenLattice}
\end{align} 
can be treated in an identical manner to the methods described in Appendix~\ref{ThetaExt} to extract $\theta_{\mathrm{eff},j}$~.
    
    \subsection{Spinor BECs driven by spin-flopping fields}
    For the data presented in Fig.~\ref{EtaExp}(b), we study an initial state with $\rho_0=0.6$, $M(0)=0$, and $\theta(0)=\pi$, which is chosen to minimize the energy dependence on $\eta$.  This initial state is a spin-coherent state defined as $\frac{1}{\sqrt{N!}} (\sum_{m_F} \sqrt{\rho_{m_F}} e^{i\phi_{m_F}} \hat{a}^{\dagger}_{m_F})^N \lvert \mathrm{vac} \rangle$, where $N$ is the total number of bosons, $N_{m_F} = \rho_{m_F} N$, $\hat{a}^{\dagger}_{m_F}$ is the creation operator for a boson in the $m_F$ hyperfine level and $\lvert \mathrm{vac} \rangle$ is the vacuum state. To imprint a desired $\eta(0)$, where $\eta(t)=\phi_{1}(t)-\phi_{-1}(t)$ is the relative phase between components of nonzero spin, we hold the atoms for a short time $t_0$ right after the initial state preparation and before the application of a pair of near resonant RF fields that drive the $m_F=0\leftrightarrow m_F=\pm1$ spin transitions. While both $\theta$ and $\eta$ evolve during the short $t_0$ duration, due to the vastly different characteristic timescales, we can tune $\eta$ effectively independent of $\theta$ using this method~\cite{Austin4}. The applied driving fields are of frequencies $2\pi(p_B\pm q)$ and are continuously applied starting at $t=0$. Here $p_B$ $(q)$ is the linear (quadratic) Zeeman shift associated with the strong static magnetic field. For typical condensates studied using this experimental sequence, the spin healing length is approximately $11~\mu\mathrm{m}$ and the Thomas-Fermi radii are approximately $(7, 7, 5)~\mu\mathrm{m}$.

\section{Extraction of $\theta$ and $c_2$}\label{ThetaExt}
    To extract $\theta$ and $c_2$ from the observed $\rho_0$ dynamics as in Fig.~\ref{FreeSpace}, we approximate $\pdv{\rho_0}{t}$ with the slope to each data point's nearest neighbors and solve Eq.~\eqref{RhoEqOfMotion} for $\sin(\theta)$. We can then use an estimated $c_2(0)$ to solve for $\theta(t_{1})$.  This allows us to approximate $\pdv{\theta}{t}$ with the slope from $\theta(0)$ to $\theta(t_{1})$ and then solve Eq.~\eqref{ThetaEqOfMotion} for $c_2(t_{1})$. Here $t_{j}$ is the holding time corresponding to the $j$-th $\rho_0$ observation. In the interaction regime, this uniquely identifies $c_2(t_{1})$ and iterating the procedure enables a $c_2(t)$ time trace to be extracted. For the free-space system analyzed in Fig.~\ref{FreeSpace}, the resulting time traces are consistent within error with a constant $c_2$ and the system can be well-described by the average value of the extracted $c_2$. These observations are consistent with the expectation that, outside of shot-to-shot variations, $c_2$ should be constant for this system. To reduce potential sources of error, we limit this method to the interaction regime. A slight variation of this technique however can allow us to determine $\theta$ for all $q$~\cite{FutureWork}.    Utilizing the average extracted $c_2$ and starting from the known initial value of $\theta(0)$, we can minimize the difference between $\pdv{\theta}{t}$ as calculated from Eq.~\eqref{ThetaEqOfMotion} and the slope from $\theta(0)$ to the potential $\theta(t_{1})$ that satisfy Eq.~\eqref{RhoEqOfMotion} to uniquely determine $\theta(t_{1})$. Iterating the process then allows a full $\theta(t)$ time trace to be extracted even in the Zeeman regime (see Fig.~\ref{FreeSpace}(d)). A similar procedure can be applied to extract the effective $\theta$ in the driven lattice system using its equations of motion (see Appendix~\ref{A}2).
    
    To reliably extract $c_2$ the experimental parameters, and in particular the initial state, must be carefully chosen because Eq.~\eqref{ThetaEqOfMotion} can diverge for some combinations of $\rho_0$ and $M$ when solved for $c_2(t)$.  For example, the extracted $c_2(t)$ is not trustable as $\rho_0$ approaches $0.5$ if $M=0$ because $1-2\rho_0$ is a factor of the coefficient of $c_2(t)$. Additionally, while the extracted $c_2(t)$ time trace has some dependence on the initial estimate of $c_2(0)$, our results indicate the extracted time traces rapidly converge with only the first handful of points displaying a significant dependence on the initial estimate of $c_2(0)$~\cite{FutureWork}. \Rev{Specifically, for the data shown in Fig.~\ref{FreeSpace}(c), $c_2$ values extracted utilizing any estimate for $c_2(0)$ such that $|c_2(0)|\gtrsim15~\mathrm{Hz}$ converge to within 1~Hz of each other by the first point shown and are practically identical for all later points.} This technique can therefore be used to robustly experimentally infer $c_2$ using a single time trace in contrast to the many time traces required to map a cross section of the phase diagram to determine $c_2$.

\section{Theoretical Treatment}\label{theory}
\subsection{Interaction Frame}
The theoretical treatment broadly follows the prior work \cite{Austin4}, which we briefly detail here as well. Starting from the Hamiltonian
\begin{align}
	\hat{H} &=\frac{c_2}{2N} \mathbf{S}^2+p_B \hat{S}_z+q\sum_i(\hat{s}_{i,z})^2 +p(t)\hat{S}_x \\
            &= \hat{H}_{S} + \hat{H}_Z + \hat{H}_D
\end{align}
where $p(t)=p(\sin(\omega_{+}t)+\sin(\omega_{-}t))$, and $\hat{H}_S$ is the interaction part of the Hamiltonian, $\hat{H}_Z$ collects the linear and quadratic Zeeman fields, and $\hat{H}_D$ refers to the time-dependent drive. We note that the full system in the laboratory additionally has an additional technical complication in that the drives and static fields point along slightly different axes, resulting in a replacement $p \rightarrow p_{\perp}$ \cite{Austin4}, which we ignore for conceptual simplicity in the following.

We first go into the interaction frame with respect to $\hat{H}_Z$ to obtain $\hat{H}' = \hat{H}'_S +\hat{H}'_{D} $ with
	\begin{align}
		\hat{H}'_S = \frac{c_2}{2N}\big[& 2 \hat{N}_0 (\hat{N}_+ + \hat{N}_-) + \hat{N}_+^2 - 2 \hat{N}_+\hat{N}_- + \hat{N}^2_- \notag\\&+ 2 (e^{i 4 \pi qt}\hat{a}^{\dagger}_+\hat{a}^{\dagger}_-  \hat{a}_0^2 + e^{-i4\pi qt} \hat{a}^{\dagger 2}_0 \hat{a}_+ \hat{a}_-)\notag\\ &+ 2N - \hat{N}_+ - \hat{N}_- \big] 
       \label{SI-eq:HI_S}
	\end{align}
and 
	\begin{align}
		\hat{H}'_{D}\!=\! - \frac{p}{2} \hat{S}_y\!+\!\frac{p}{2\sqrt{2}} \left(  i e^{-4\pi i q t} \hat{a}_0 \hat{a}_{+}^{\dagger}\!+\!i e^{4\pi i q t} \hat{a}_0^{\dagger} \hat{a}_{-} \!+\!\mathrm{h.c.}\right), 
        \label{SI-eq:HI_D}
	\end{align}
where we dropped terms rotating with $p_B$, but retained $q$-dependent terms, consistent with the hierarchy $p_B \gg q$. %

In this interaction frame the part of the original interactions corresponding to the spin-flip process $00 \rightarrow +-$ become time-dependent due to the energy cost of $2 q$ required for this conversion in presence of the quadratic Zeeman shift, whereas the pure density terms remain time-independent. Additionally, the original drive resonant with the $0\rightarrow +$ and $0 \rightarrow -$ transitions is seen to result in a time-independent $\hat{S}_y$ drive term, as well as time-dependent transition terms ($2 q$ periodic terms). 

The time dynamics presented in Fig.~\ref{Eta} are based on exact time-evolution under the Hamiltonian $\hat{H}'$ for $N=140$ particles.
\subsection{Floquet Hamiltonian}
As a time-periodically driven Hamiltonian with period $T=2 \pi/\Omega$ and frequency $\Omega= 2 q/\hbar$,  the above Hamiltonian can be analyzed within the framework of Floquet theory. This allows to use effective time-independent Hamiltonian $\hat{H}_F$ which can be expanded in powers of the inverse frequency \cite{Bukov2015, Eckardt2015,Kuwahara2016} to describe the physics of the model.

In this work we use the Floquet-Magnus expansion at second order, $\hat{H}_F \approx \hat{H}_F^{(0)} + \hat{H}_F^{(1)} + \hat{H}_F^{(2)}$. Here $\hat{H}_F^{(0)}$ is the average static Hamiltonian, and the Floquet correction terms are respectively given by
\begin{equation}
    \hat{H}_F^{(1)} = \frac{1}{2 T i \hbar} \int_{0}^{T} dt_1 \int_{0}^{t_1} dt_2 [H(t_1), H(t_2)] 
\end{equation}
and
\begin{align}
    \hat{H}_F^{(2)} = \frac{1}{6 T (i \hbar)^2}\!\int_{0}^{T}\!\!\!dt_1\!\int_{0}^{t_1}\!\!\!dt_2\!\int_{0}^{t_2}\!\!\!&dt_3 \, [H(t_1), [ H(t_2),H(t_3)]] \notag\\&+ [H(t_3), [ H(t_2),H(t_1)]] 
\end{align}
The spectral properties and thermal micro-canonical thermal expectation values presented in Fig.~\ref{Eta} \Rev{and Fig.~\ref{EtaExp}} are based on this effective Floquet Hamiltonian.

\Rev{We note that generically interacting periodically driven quantum systems would be expected to heat up to infinite temperature. In the high frequency limit this heating process is expected to occur on exponentially or parametrically long times. We note that this work operates in an intermediate driving regime, where these general expectations may not necessarily apply. However, our simulations of the exact time dynamics under Eqs.~\eqref{SI-eq:HI_S}-\eqref{SI-eq:HI_D} do not show signatures of Floquet heating for the time-scales considered here.} 
\vspace{1cm}
   
\bibliography{References}

\end{document}